\documentstyle [12pt] {report}
\begin{document}
\setlength{\baselineskip}{24pt}

\title{\large \bf  $R + R^2$ GRAVITY AS $R +$ BACKREACTION\thanks
{Accepted for Publication in Phys. Rev. D}}

\author{\bf J. H. Kung \\
(jkung@abacus.bates.edu)\\
Department of Physics \\
Bates College\\
44 Campus Avenue\\
Lewiston, Maine 04210}

\date{}
\maketitle
\begin{abstract}

Quadratic theory of gravity is a complicated constraint system.  We
investigate some consequences of treating quadratic terms
perturbatively (higher derivative version of backreaction effects),
which is consistent with the way existence of quadratic terms was
originally established (radiative loop effects and renormalization
procedures which induced quadratic terms).  We show that this
approach overcomes some well known problems associated with
higher derivative theories, i.e.,  the physical gravitational degree of
freedom remains unchanged from those of Einstein gravity.

Using such an approach, we first study the classical cosmology of
$R + \beta R^2$  theory coupled to matter
 with a characteristic
$\rho \propto a(t)^{-n}$ dependence on the scale factor. We show
that for
$n> 4 \ \ (i.e., p > {1\over 3}\rho)$ and for a particular sign of
$\beta$, corresponding to non-tachyon case,  there is no big bang in
the traditional sense.  And therefore, a contracting FRW universe
($k>0, k=0,k<0$) will rebounce to an expansion phase without a total
gravitational collapse.

We then quantize the corresponding mini-superspace model that
resulted from treating the $\beta R^2$ as a perturbation.  We
conclude that the potential $W(a)$, in the Wheeler DeWitt equation
$\left[-{\partial^2\over \partial a^2} + 2W(a)\right]\psi (a) =0 $,
develops a repulsive barrier near $a \approx 0$ again for
$n> 4 \ \ (i.e., p > {1\over 3}\rho)$ and for the sign of $\beta$ that
corresponds to non-tachyon case.

sign of $\beta$.%

Since $a \approx 0$  is a classically forbidden region, the probability
of finding a universe with singularity ($a=0$) is exponentially
suppressed. Unlike quantum cosmology of Einstein's gravity, the
formalism has dictated an appropriate boundary (initial) condition.

Classical and quantum analysis demonstrate that a minimum radius
of collapse increases for a larger value of  $\vert \beta \vert$.

It is also shown that, to first order in $\beta$, $\beta R^2$ term has
no effect during the radiation ($p = {1\over 3}\rho$) and inflationary
($p = -\rho$) era. Therefore, a deSitter phase can be readily
generated by incorporating a scalar field.

\end{abstract}

\begin{flushleft}{\large \bf I. INTRODUCTION}
\end{flushleft}
\setcounter{chapter}{1}
\setcounter{equation}{0}

Since the discovery of the singularity theorem of Hawking and
Penrose [1], the speculation of creating non singular theory by
incorporating quantum property of gravity and/or using
 non-Einstein gravity has attracted some interest.  Because of
advancements in quantum field theory,  the two avenues of
speculations seem to be mathematically related.  That is,  even if one
starts with Einstein's gravity, renormalization consideration dictates
that action for gravity must have terms that are quadratic in Ricci
tensor [2].

In quantum cosmology, canonical quantization is the preferred
formalism.  This is because, for a universe that is homogeneous and
isotropic in large scales, there is no asymptotically ``in'' and ``out''
fields, which is necessary in order to implement covariant
quantization formalism.  The task of identifying dynamical degrees of
freedom for quadratic gravity has been reduced to solving constraint
system by Boulware [3].  Because of the technical difficulties of
solving such constraints, quantum cosmology for quadratic gravity
has been solved for only simple systems like vacuum [4,5].

those

In our paper, we explore consequences of viewing $R+R^2$ gravity as
$R +$ perturbation (higher derivative version of backreaction) [6]. In
essence, in this approach, the physical gravitational degree of
freedom is not changed from that of Einstein gravity.  It is shown
that this view overcomes the bulk of technical difficulties with the
higher derivative content of quadratic gravity.  In section (II), we
argue that this view is also in agreement with the way the existence
of quadratic terms were originally established (via renormalization
procedures that treats quadratic terms perturbatively).

Using such an approach (section III), we study the classical
cosmology of $R+\beta R^2$ theory coupled
to matter that has a characteristic
$\rho \propto a(t)^{-n}$ dependence on the scale factor. We show
that for $n> 4 \ \ (i.e., p > {1\over 3}\rho)$ and for a particular sign
of $\beta$, corresponding to non-tachyon case,  there is no big bang
in the traditional sense.
  And therefore, even for a close FRW metric, the universe will
rebounce without a complete collapse.

In section (IV), we quantize the corresponding mini-superspace
model, which resulted from treating the $\beta R^2$ as a
backreaction.  We conclude that the potential $W(a)$, in the Wheeler
DeWitt equation
 $\left[-{\partial^2\over \partial a^2} + 2W(a)\right]\psi (a) =0 $,
develops a repulsive barrier near $a \approx 0$ again for
$n> 4 \ \ (i.e., p > {1\over 3}\rho)$ and for a particular sign of
$\beta$, which corresponding to non-tachyon case


Sign conventions used in this paper are as follows. $g=(-,+,+,+)$,
$R_{\mu\nu} -{1\over 2}g_{\mu,\nu}R = \left (+ \right )8\pi
GT_{\mu\nu}$, $\left (+ \right ) R(\mu,\nu) =
\nabla_{\mu}\nabla_{\nu}- \nabla_{\nu}\nabla_{\mu}-
\nabla_{\left[\mu,\nu\right]}$.

\begin{flushleft}{\large \bf II. PRELIMINARIES}
\end{flushleft}
\setcounter{chapter}{2}
\setcounter{equation}{0}
The most general quadratic action for gravity coupled to matter is

\begin{equation} I = -{1\over 16\pi G}\int d^4x\sqrt{-g}R -\int d^4x
\left
[\beta_1R^2+\beta_2R_{ab}R^{ab}+\beta_3R_{abcd}R^{abcd}\right] +
I_{matter} + {\rm surface \; term}\end{equation}

We have formally included a surface term to cancel any boundary
term that would result in applying the variational principle.  By
dimensional analysis, $\beta_1, \beta_2, \beta_3$ are dimensionless.
We will be interested in applying the formalism to homogeneous and
isotropic metric, i.e., Wely tensor vanishes $C_{abcd}=0$[2].
By definition of Weyl tensor, $C_{abcd}C^{abcd}=R_{abcd}R^{abcd}
-2R_{ab}R^{ab} +{1\over 3}R^2$.
This gives one relationship among the possible quadratic terms.

The second relationship is from the 4 dimensional generalization of
Gauss Bonnet formula [2],

\begin{equation} R^2 - 4R_{ab}R^{ab} + R_{abcd}R^{abcd} = {\rm exact
\; derivative.}\end{equation}

The two relationships, combined with the fact that Euler Lagrange
equations are unchanged by addition of an
exact differential, allow any two of
$\beta_1, \beta_2, \beta_3$ to be set equal to zero in the action
(2.1).  We choose to set $\beta_3=\beta_2=0$.

Upon variation of the metrics, the resulting Euler Lagrange equations
are
\begin{equation} {1\over2}Rg_{ab} - R_{ab}+16\pi
G\beta\left({1\over 2}R^2g_{ab} - 2RR_{ab} +
2R_{;\sigma}^{;\sigma}g_{ab} - 2R_{;a;b}\right) = 8\pi
GT_{ab}.\end{equation}

The trace of this equation is
\begin{equation} 6\cdot 16\pi G\beta R_{;\sigma}^{;\sigma} + R =
8\pi GT,\end{equation}

which reduce to a familiar forms for $\beta =0$.

 For $\beta =0$ case, it is well known that both the left and the right
side of (2.3) vanish under covariant derivative (the right hand side
by local conservation of energy momentum tensor and the left hand
side by Bianchi identity).  It is interesting to note that this is true
even for $\beta \neq 0$ (the additional term is also covariant
constant by the virtue of Bianchi Identity).  This consistency is
expected since local conservation of energy momentum tensor and
Bianchi identity are intimated related to reparametrization
invariance of the action (2.1).

In this paper, we shall consider only the simplest metric, that of
spatially homogeneous and isotropic universe (i.e., FRW metric)

\begin{equation} ds^2 = -dt^2 + a^2\left ( {dr^2\over {1-kr^2}} +
r^2d\Omega_2^2\right ),\end{equation}

with the standard energy momentum tensor
\begin{equation}T_{ab} = pg_{ab} + (p + \rho)U_aU_b\\
\;\;\;\; U^0=1, \;\; U^i=0, i=1,2,3. \end{equation}

Using,

\begin{equation}R = -6\left( {\ddot a\over a} +
\left ( {\dot a\over a}\right)^2 + {k\over a^2} \right),\end{equation}

for Ricci scalar, one gets for the time-time component of (2.3).
\begin{equation} \left({\dot a \over a}\right)^2 + {k\over a^2} -
{16\pi G\beta\over 3}\left ( {1\over2}R^2 + 6R{\ddot a\over a}
-2{\partial_t \left( a^3\partial_tR\right)\over a^3} +
2\partial_t^2R\right) = {8\pi G\rho\over 3},\end{equation}

and for any one of space-space components,
\begin{equation} 2{\ddot a\over a} + \left({\dot a \over a}\right)^2 +
{k\over a^2} - 16\pi G\beta\left ( {1\over2}R^2 + 2R\left( {\ddot
a\over a} +2\left({\dot a\over a}\right)^2 + {2k\over a^2}\right) - 2
{\partial_t\left(a^2\partial_tR\right)\over a^2}\right) = -8\pi
Gp.\end{equation}

As is evident, for $\beta \neq 0$, (2.8) is transformed from first
order to third order, and (2.9) is transformed from second order to
fourth order differential equation.

Even on a classical level, there are pathological problems with higher
derivative theories [6,7].  One of which is the need for additional
initial
conditions to completely specify a system, and whether solutions
obtained by solving (2.8 -2.9) for $\beta \neq 0$ have well behaved
properperties as $\beta \rightarrow 0$, and existence of run-away
solutions.

A review of motivation for studying quadratic gravity
(i.e., renormalization consideration) seems to offer an alternate
method of handling higher derivative theories of gravity.

In perturbative covariant quantization, even if one starts with the
Einstein's action coupled to matter as the bare action,

\begin{equation} I_{bare} = \int\sqrt{-g}R +
I_{matter}\end{equation}

by quantum loop effects from both gravity and matter, the effective
action acquires terms with higher derivatives

\begin{equation} I_{effective} = I_{bare} + \int\left[ \alpha_1R^2 +
\alpha_2R_{ab}R^{ab} +
\alpha_3R_{abcd}R^{abcd}\right]\end{equation}

with divergent $\alpha_1, \alpha_2, \alpha_3.$  The precise nature
of divergence depends on a choice for a matter field.

The perturbative renormalization prescription is to add terms to the
bare action to precisely cancel these infinities, i.e.,
\begin{equation} I_{bare} \rightarrow \int\sqrt{-g}R + I_{matter} +
I_{counter-term}\end{equation}

\begin{equation} I_{counter-term} =  \int\left[ (\beta_1-
\alpha_1)R^2 + (\beta_2-\alpha_2)R_{ab}R^{ab} + (\beta_3-
\alpha_3)R_{abcd}R^{abcd}\right].\end{equation}

This renders the resulting effective action finite, which may be used
for semiclassical analysis. (For a fuller discussion of unitarity and
renormalizability, please see [2,7].)  The crucial observation is that
physical degree of freedom for asymptotic "in" and "out" fields were
those of Einstein action.  Moreover, even when the bare action had
higher derivative counter-terms (2.12), the higher derivative terms
were treated perturbatively and not on an equal footing with the
Einstein's action.  This is in mark contrast with [3]. The details of our
proposed procedures will be shown in section III.   Our proposal for
gravity is
similar to Jaen's et al.[6] procedure for reducing a general Lagrangian
system with
arbitrary higher derivatives into a second-order differential system.

A further support of this interpretation of theories in which higher
derivative were induced can be found in classical electrodynamics.
For a pedagogical review, please see Barut[8].  Consider a non
relativistic harmonic oscillator.  It obeys the Newton's law,

\begin{equation}m{\bf {\ddot R}} = {\bf {F_{ext}}}= -m\omega^2{\bf
R}. \end{equation}

If the particle is also charged, than the accelerated particle emits
radiation and in turn must effect change on the motion of particle.
One can take into account of this radiative loss of energy by an
effective
radiative backreaction force

\begin{equation}m{\bf {\ddot R}} = {\bf {F_{ext}}}+{\bf {F_{rad}}}
\end{equation}

with
\begin{equation}{\bf {F_{rad}}} = {2e^2\over 3 c^3}{\bf
\stackrel{\ldots}{R}}\end{equation}

The resulting equation of motion is changed from second to third
order.  Besides the necessity of additional initial conditions,  (2.15)
also has an unphysical run away solution. For example, if
 ${\bf {F_{ext}}} =0$, than there should be no acceleration and hence
no radiative loss, i.e., ${\bf {\ddot R}}  =0$.  Yet, it is straightforward
to verify that (2.15) admits an unphysical solution ${\bf {\ddot R}}
={\bf C}e^{t\over \tau}$ with $\tau = {2e^2\over 3 mc^3}$.

For a weak radiative loss, there are well known approximate
methods of handling these problems, which result in physically
acceptable solutions.  For
SHO problem, weak radiative loss means that the solution should be
harmonic to
first order, ${\bf \stackrel{\ldots}{R}} \approx -\omega^2 {\bf  {\dot
R}}$.  Upon
substitution, the resulting effective equation of motion is returned to
the original second order,
i.e.,

\begin{equation}{\bf {\ddot R}}  +\omega^2\tau{\bf {\dot R}} +
\omega^2{\bf R}=0.\end{equation}

For a general ${\bf {F_{ext}}}$, one can still eliminate unphysical
solutions by using an
integration factor and surgically choosing initial condition for
${\bf {\ddot R}}(0)$.In either case, the lesson is that the induced
higher derivative forces
were treated as backreactions which did not increase the physical
degree of freedom.

The analogy is even stronger for action at a distant treatment of
classical
electrodynamics.  In this case, the backreaction force is not deduced
by balancing energy but by a dynamical process.  Here the
backreaction force can be split into contributions from near and far
field produced by the particle.  The backreaction force from the far
field is relativistic generalization of (2.16), and backreaction force
from
the near field results in classical mass renormalization.

The similarity between induced higher derivative (via radiative
processes) in classical
electrodynamic and the present problem with gravity is obvious.
 Therefore, using these as motivations and possibly even as
justifications,  we shall also treat the higher derivative terms
perturbatively when implementing the canonical quantization
procedure (i.e., Wheeler DeWitt equation).  The rest of the paper can
be categorized as consequeces of such an approach.

\begin{flushleft}{\large \bf III. CLASSICAL EVOLUTION OF $R + R^2$
GRAVITY COUPLED TO MATTER}
\end{flushleft}
\setcounter{chapter}{3}
\setcounter{equation}{0}

In (2.3-2.4), we are interested in treating contributions from $R^2$ as
perturbation.  We will use $\beta$, as a dimensionless expansion
parameter and study the first order correction to equation of motion.
The value of $\beta$ is of course unknown, but in order to
implement perturbation method, we will also have to assume that
$\beta$ is small.  Hopefully, results obtained by assuming small
$\beta$ will only be amplified
by a larger $\beta$.  We will return to this issue at the end.

By (2.4), $R=8\pi GT + O(\beta)$ and
$R_{ab} = -8\pi G\left( T_{ab} -{1\over 2}Tg_{ab}\right) + O(\beta) $

Therefore, Eq (2.3) is
\begin{equation} {1\over2}Rg_{ab} - R_{ab} = \tilde GT_{ab} +2{\tilde
G}^2\beta\left({1\over 2}{\tilde G} T^2g_{ab} - 2{\tilde G}TT_{ab} -
2T_{;\sigma}^{;\sigma}g_{ab} + 2T_{;a;b}\right)  +
O(\beta^2).\end{equation}

We have introduced a notation $\tilde G \equiv  8\pi G$.

Using (2.5-2.7), to first order in $\beta$,  the time-time component is

\begin{equation} \left({\dot a \over a}\right)^2 + {k\over a^2} =
{1\over 3}{\tilde G}\rho - 2{\tilde G}^2\beta\left ( {1\over 2}{\tilde
G}(3p - \rho)(p + \rho) + 2{\dot a\over a}\partial_t(3p - \rho)
\right). \end{equation}

And any one of the space-space components gives the single
equation

\begin{equation} 2{\ddot a\over a} + \left({\dot a \over a}\right)^2 +
{k\over a^2} = -{\tilde G}p + {\tilde G}^2\beta\left({\tilde G}(3p -
\rho)(p + \rho) -{ 4\partial_t(a^2\partial_t(3p - \rho))\over
a^2}\right). \end{equation}

The matter sector must satisfy local conservation of energy
momentum tensor
\begin{equation} {d(\rho a^3)\over da} = -3pa^2.\end{equation}

For $\beta =0$, it is well known that solutions obtained by (3.2)  and
(3.4) automatically satisfy (3.3).  As pointed out in section (II), this is
guaranteed even for $\beta \neq 0$ by reparametrization invariance
of the action (2.1).  Therefore, we shall proceed to solving (3.2) with
(3.4).

{}From the form of (3.2), one can immediately extract several
conclusions.  First, to first order in $\beta$, the $R^2$ has no
contribution in the radiation era because of
$p_{rad} = -{1\over 3}\rho_{rad}$ .  Second, if energy momentum
tensor ever becomes dominated by an almost constant potential of a
scalar field, ($-p_{\phi} \approx \rho_{\phi} \approx  V(\phi)
\approx constant$) then we can again conclude that the contribution
from $R^2$ vanishes (to first order in $\beta$).  Therefore, a possible
deSitter phase in $R + \beta R^2$ with a scalar field $\phi$ should be
more or less identical with a deSitter phase in standard Einstein
gravity.  It is of course a separate question whether one can generate
an inflationary phase in $R+R^2$ gravity without fine-tuning.

In comparison, Mijic[9] and Page[10] have studied the large $\beta$
range and concluded that, even for a vacuum, gravity alone can
generate an inflationary phase in $R+R^2$ gravity.

Now let us assume that during any epoch in the evolution of
universe, the universe  is dominated by a matter with a
characteristic dependence on the scale factor
(i.e.,  $\rho = {\rho_o\over a^n}$), n=3 for matter era, n=4 for
radiation era, and n=0 for an inflationary era, etc.   From
conservation
of energy and momentum tensor (3.4), we get $p={n-3\over
3}{\rho_o\over a^n}.$

Therefore

\begin{equation} (3p - \rho)(p + \rho) = {1\over 3}n(n-
4){\rho_o}^2\end{equation}

and

\begin{equation} {\dot a\over a}\partial_t(3p - \rho) = -n(n-
4)\left({\dot a\over a}\right)^2{\rho_o\over a^n} = -n(n-4)\left(-
{k\over a^2} + {1\over 3} {\tilde G}\rho\right){\rho_o\over a^n} +
O(\beta).\end{equation}

After simplifying, we get

\begin{equation} {\dot a}^2 + 2U(a) = 0 \end{equation}

\begin{equation} 2U(a) = k \left( 1 + 4n(n-4){\tilde G}^2\beta
{\rho_o\over a^n}\right) - {1\over 3}{\tilde G}{\rho_oa^2\over
a^n}\left( 1 + 3n(n-4){\tilde G}^2\beta {\rho_o\over a^n}\right)  +
O(\beta^2)\end{equation}

which can be interpreted as an equation describing a particle with a
unit mass in a potential $U(a)$.  For $\beta = 0$, the form of  $U(a)$
is shown in Figure 1.  As expected, depending on the sign of
curvature of three space,  ($k>0, k<0, k=0$) the universe evolves
either as a bound state, unbounded state, or as a critically opened
state, respectively.

Now, for $\beta \neq 0$, the contributions from $R^2$ depend
crucially on the sign of $\beta$.  First, we note that $\beta < 0$
corresponds to non-tachyon case.  This can be readily deduced from
(2.4).  As noted by [3-6], $\phi \equiv R$ evolves like a scalar field
with $m^2 = {-1\over 6\cdot 16 \pi G\beta}$.  Therefore, $\beta < 0$
is needed to eliminate tachyons. Mijec et al.[9], Stelle[11],
Teyssandier
and Tourrenc[12], Barrow and Ottewill[13], Mazzitelli and
Rodrigues[14]
have noted that $\beta < 0$ is necessary otherwise the Hubble
parameter grows without bound.
{}.

Confining ourselves to $\beta < 0$, notice that for
$ n > 4,  (i.e., p > {1\over 3}\rho)$,  $U(a)$ develops a potential
 barrier near $a \approx 0$ (Figure 2).  The interpretation is
straightforward.
For such a case, a contracting FRW universe ($k>0, k=0,k<0$) will
rebounce to an expansion phase without a total gravitational collapse.
The graph of region $ p > {1\over 3}\rho$ is shown in
Figure 3.a.  For comparison, the strong energy condition for isotropic
and homogeneous system [15], which predicts an existence of
singularity, gives $ \rho + 3p \geq 0$ and $\rho + p \geq 0$, Figure
3.b. As is obvious from Figure 3.a + 3.b,  most of the parameter space
$(\rho, p)$ which is predicted to have singularity in Einstein gravity
do not have singularity in $R + \beta R^2$ gravity.

Several comments are in order.  First, strong energy condition comes
from study of Raychaudhuri's equation, which describes how a
congruence of timelike geodesices deviate from one another.  Indeed,
the appropriate strong energy conditions for $R+R^2$ is different
from that of Einstein gravity and can be shown to be in agreement
with the present result [16].

Second,  Page[10] and Coule + Madsen[17] have studied a vacuum
model
and noted a bounce solution for only $k<0$.  The difference with the
present work is more than a vacuum versus a nonvacuum model.


As explained in detail in section II, in our approach, we have treat
the $ \beta R^2$ as a backreaction on Einstein gravity [6].  In essence,
the physical gravitational degree of freedom has not been changed
from that of Einstein gravity.  This approach has the advantage of
having a smooth limit as $\beta \rightarrow 0$, and avoids the
pathological situation with the  necessity for ``extra" initial conditions
in higher derivative theories.

On the other hand, [10,17] method is straightforward yet the field
content, gravitational degree of freedom, is different from that of
Einstein gravity, and the existence of limit as $\beta \rightarrow 0$
is questionable.

\begin{flushleft}{\large \bf IV. QUANTIZATION OF MINISUPERSPACE
MODEL}
\end{flushleft}
\setcounter{chapter}{4}
\setcounter{equation}{0}

The prediction from classical analysis is indeed interesting, but near
the Planck time, a quantum analysis is needed.  There have been
some literature on quantizing mini-superspace model for $R + R^2$
gravity [4,5].  In the literature, because of technical difficulties of
quantizing higher derivative theories, only very simple cases
(i.e. no coupling to matter) have been considered.  With our proposal
of treating $R^2$ term as a perturbation, we can readily address
more realistic systems with matter.

Even though $R^2$ contains terms with higher time derivative than
in the Einstein's action, we are treating it as a perturbation.
Therefore, the physical degrees of freedom (e.g., canonical momenta)
are determined by Einstein's action alone [6].

 For FRW metric, the only gravitational degree of freedom is the scale
factor a(t).  Therefore, canonical momentum conjugate to a(t) is [18]

\begin{equation} \pi_a \equiv {\delta I_G\over \delta{\dot a}} =
-{3 V_3\over 4\pi G k^{3/2}}a\dot a = -{3 \pi \over 2 G
k^{3/2}}a\dot a .
\end{equation}

In terms of $\pi_{a}$, the time-time component of (3.2)  is

\begin{equation} {\pi_a}^2 + 2W(a) = 0 \end{equation}

\begin{equation} 2W(a) \equiv \left({12\pi^2\over {\tilde
G}}\right)^2{2U(a)a^2\over k^3}.\end{equation}

$U(a)$ was defined in (3.8).

We will not need the expression for Wheeler-DeWitt equation in its
most general form.  To quantize the system, we replace
$\pi_a \rightarrow -i{\partial\over \partial a}$ in (4.2) to get

\begin{equation} \left[ -{\partial^2\over \partial a^2} +
2W(a)\right]\psi (a) = 0. \end{equation}

Because we are interested in semiclassical analysis, we have
neglected the factor ordering parameter [19-21].  A comment is in
order.  Rigorously, there should also a "kinetic" term
${\partial^2\over \partial {\phi}^2}$ representing quantum
fluctuation of some matter field.  Indeed, the analysis gets quite
involved and it will be addressed in subsequent work. Here, we shall
be satisfied with particular ``semiclassical" analysis in which only the
gravitational sector is treated quantum mechanically [18, 22-25]

As before, we will assume that during any epoch in the evolution of
universe, the universe is dominated by a single matter with
 $\rho = {\rho_o\over a^n}$.  For example, scalar field conformally
coupled to gravity would be n=4, a massive quantum field would be
n=3, and during a possible inflationary era $\rho_{\phi} \approx
V(\phi)
\approx constant$ or n=0, etc.

  For $\beta = 0$  and closed ($k > 0$),  the resulting Schrodinger
equation resembles quantum mechanical description of unit mass in
a bound state potential $W(a)$ [18,22-24] (Figure 4). The form of
$W(a)$ for a
universe which has undergone a standard inflationary
phase (via a scalar field) is shown in Figure 5.
Since
$a=0$ is boundary of a physically allowed region, a boundary
condition (initial condition)  $\psi(a=0)$ must be specified to
completely describe a system [26-29].

Now, for $\beta \neq 0$, the situation is similar to what we have
discovered by classical analysis.  Notice that for again $\beta < 0$ and
for $n> 4 \ \ (i.e., p > {1\over 3}\rho)$ $W(a)$ develops a potential
barrier
 near $a \approx 0$ (Figure 6).

region.

Physical interpretation of $\psi$  is unclear in quantum cosmology,
but if one may interpret ($\psi \propto Exp(-\vert \;\;\vert)$) as an
indication of small probability, then our analysis indicated that a
model with  $p > {1\over 3}\rho$ and $\beta < 0$ will have a very
small probability of a big bang or total recollapse. Notice that unlike
Einstein's gravity, there is no issue of boundary condition. There is a
shortcoming.  In Figure 6, the dotted region is precisely when the
first order approximation breaks down.

\begin{flushleft}{\large \bf VII. CONCLUSION}
\end{flushleft}
\setcounter{chapter}{7}
\setcounter{equation}{0}

In this paper, we have studied $R + \beta R^2$ by treating $R^2$
term as a perturbation.  For FRW metric with matter, using classical
analysis, we have shown that for  $n> 4 \ \ (i.e., p > {1\over 3}\rho)$
and for a particular sign of $\beta $ that corresponds to non-tachyon
case, there is no big bang in the traditional sense.  And therefore, a
recollapsing FRW closed universe will rebounce without a complete
collapse.

{}From quantization of corresponding minisuperspace model, we have
shown that the potential $W(a)$, in the Wheeler DeWitt equation
$\left[-{\partial^2\over \partial a^2} + 2W(a)\right]\psi (a) =0 $,
develops a repulsive barrier near $a \approx 0$ again for  $n> 4 \ \
(i.e., p > {1\over 3}\rho)$ and for the sign of $\beta $ that
corresponds to non-tachyon case.
Since $a \approx 0$ is now strictly a classically forbidden region, a
probability of finding a universe with singularity ($a=0$) is
exponentially suppressed.

And in closing, we can address the effects of a larger value of
$\beta$.  From (3.7-3.8) + Figure 2 and (4.2-4.3) + Figure 6,  the
minimal classical radius of collapse and the size of classically
forbidden region increase with larger $\vert\beta\vert$,
respectively.

\begin{center}{\large \bf IX. ACKNOWLEDGMENTS}
\end{center}
\setcounter{chapter}{9}
\setcounter{equation}{0}
I would like to thank Mr. Paul Riley for kindly proof reading the
manuscript.

\newpage

\begin{center}
{\large \bf FIGURE CAPTIONS}
\end{center}
\begin{enumerate}
\item $U(a)$ for $k>0$ and $k<0$ universe, $\beta =0.$.
\item $U(a)$ $ n > 4, \beta <0.$
\item Graph of region for $p > {1\over 3}\rho$, Figure 3.a.  Graph of
regions $p \geq -{1\over 3}\rho$ and $p \geq -\rho$, for strong
energy condition, Figure 3.b.
\item $W(a)$ for $k>0$ universe, $\beta =0.$
\item $W(a)$ for very early universe during inflationary era, $n
\approx 0$, $\beta =0$
\item $W(a)$ for $ n > 4, \beta <0.$

\end{enumerate}

\newpage
\begin{center}
{\large \bf REFERENCES}
\end{center}
\begin{enumerate}
\item S.W. Hawking and G. F. R. Ellis, {\it The Large Scale Structure of
Space-Time} (Cambridge University Press, 1973).

\item S. Deser, in {\it Quantum Gravity}, edited by C. J. Isham, R.
Penrose and D. W. Sciama, (Oxford University Press, 1975).

\item D. Boulware, in {\it Quantum Theory of Gravity}, edited by S.
Christensen (Adam Hilger, 1984).

\item S. W. Hawking and J. C. Luttrell, Nucl. Phys. {\bf B 247}, 251
(1984).

\item U. Kasper, Class. Quantum Grav. {\bf 10}, 869 (1993).

\item X. Jaen et al., Phys. Rev. {\bf D 34}, 2302 (1986).

\item A. Strominger, in {\it Quantum Theory of Gravity}, edited by S.
Christensen (Adam Hilger, 1984).

\item A. Barut, {\it Electrodynamics and Classical Theory of Fields
and Particles} (Dover Publications Inc., 1980).

\item M. Mijic et al., Phys. Rev. {\bf D 34}, 2934 (1986).

\item D. Page, Phys. Rev. {\bf D 36}, 1607 (1987).

\item K. Stelle, Gen. Relativ. Gravit. {\bf 9}, 353 (1978).

\item P. Teyssandier and P. Tourrenc, J. Math. Phys. {\bf 24}, 2793
(1983).

\item J. Barrow and A. Ottewill, J. Phys. A: Math. Gen. {\bf 16}, 2757
(1983).

\item F. Mazzitelli and L. Rodrigues, Phys. Lett. {\bf B 251}, 45
(1990).

\item R. Wald, {\it General Relativity} (University of Chicago Press,
1984).

\item J.H. Kung, in preparation.

\item D. Coule and M. Madsen, Phys. Lett. {\bf B 226}, 31 (1989).

\item J. H. Kung, Gen. Rel. Grav. {\bf 27}, 35, (1995).

\item J. J. Halliwell, Phys. Rev. {\bf D38}, 2468 (1988).

\item C.W. Misner, in {\it Relativity}, edited by M. Carmeli, S. I.
Fickler, and L. Witten, (Plenum, New York, 1970).

\item S.W. Hawking and D. N. Page, Nucl. Phys. {\bf B 264}, 185
(1986).

\item B. S. DeWitt, Phys. Rev. 160, 1113 (1967).

\item J. Narlikar and T. Padmanabhan, {\it Gravity, Gauge Theories
and Quantum Cosmology} (D. Reidel Publishing, 1986).

\item M. Cavaglia, V. de Alfaro, and A. Filippov, Int. Jou. Mod. Phys.
{\bf A10}, n.5 (1995).

\item Y. Peleg, Brandeis Preprint, BRX-TX-342.

\item J. B. Hartle and S. W. Hawking, Phys. Rev. {\bf D28}, 2960
(1983).

\item S. W. Hawking, Nucl. Phys. {\bf B 239}, 257 (1984).

\item A. Vilenkin, Phys. Rev. {\bf D 33}, 3560 (1986).

\item A. Vilenkin, Phys. Rev. {\bf D 37}, 888 (1988).

\end{enumerate}

\end{document}